\documentclass[11pt]{article}

\usepackage[english]{babel}
\usepackage[utf8]{inputenc}
\usepackage{natbib}

\usepackage{algorithm} 
\usepackage{algorithmic}

\usepackage{multirow}
\usepackage{colortbl}
\usepackage{array,booktabs}
\usepackage{xcolor}
\usepackage{makecell}
\usepackage{pifont}

\makeatletter
\def\BState{\State\hskip-\ALG@thistlm}
\makeatother

\usepackage{amssymb, authblk}
\usepackage{amsmath, bbm}
\usepackage{fullpage}
\usepackage{amsthm, natbib, hyperref}
\usepackage{graphicx}
\usepackage{verbatim}

\usepackage{threeparttable}
\usepackage{algorithm}

\usepackage[usestackEOL]{stackengine}

\DeclareMathAlphabet{\mathpzc}{OT1}{pzc}{m}{it}
\usepackage{cleveref}
\usepackage{hyperref}[]
\hypersetup{
	colorlinks=true,
	linkcolor=blue,
	filecolor=magenta,      
	urlcolor=cyan,
	citecolor=blue,
}

\allowdisplaybreaks

\DeclareMathOperator*{\argmin}{arg\,min}

\usepackage{xspace}

\title{Glucodensities: a new representation  of   glucose profiles  using distributional data analysis}

\author{Marcos Matabuena$^{1,2,*}$, Alexander Petersen$^{3}$, Juan C.Vidal$^{2}$ and Francisco Gude$^{1}$ \\
	$^{1}$Unidad de Epidemiolog\'{i}a Cl\'{i}nica, Hospital C\'{i}nico Universitario de  Santiago de Compostela, Spain\\ $^{2}$CiTIUS (Centro Singular de Investigaci\'{o}n en Tecnolox\'{i}as Intelixentes), Universidade de Santiago de Compostela, Spain \\  $^{3}$Department of Statistics and Applied Probability, University of California, Santa Barbara \\
	$^{*}$\url{marcos.matabuena@usc.es}
	\\}

\date{\today}

\begin{document}
\maketitle
\begin{abstract}
	
Biosensor data has the potential ability to improve disease control and detection. However, the analysis of these data under free-living conditions is not feasible with current statistical techniques. To address this challenge, we introduce a new functional representation of biosensor data, termed the glucodensity, together with a data analysis framework based on distances between them. The new data analysis procedure is illustrated through an application in diabetes with continuous-time glucose monitoring (CGM) data. In this domain, we show marked improvement with respect to state of the art analysis methods. In particular, our findings demonstrate that i) the glucodensity possesses an extraordinary clinical sensitivity to capture the typical biomarkers used in the standard clinical practice in diabetes, ii) previous biomarkers cannot accurately predict glucodensity, so that the latter is a richer source of information, and iii) the glucodensity is a natural generalization of the time in range metric, this being the gold standard in the handling of CGM data. Furthermore, the new method overcomes many of the drawbacks of time in range metrics, and provides deeper insight into assessing glucose metabolism.  
	
	
\end{abstract}

\section{Introduction}

The steadily increasing availability and prominence of biosensor data have given rise to new methodological challenges for their statistical analysis. A primary feature of these data is that the monitored individuals are in free-living conditions, making a direct analysis of the recorded time series between groups of patients problematic if not infeasible.  A clear example of such data is found in the study of diabetes, where continuous glucose monitoring (CGM) is increasingly used. The elevation of glucose is distinct between individuals and is influenced by factors such as mealtimes, diet composition, or physical exercise  \citep{doi:10.1177/0962280214520732}.  Consequently, an exciting topic of debate is how to exploit the enormous wealth of information recorded by CGM to draw more reliable conclusions about the glucose homeostasis rather than the cursory summary measures such as fasting plasma glucose (FPG) or glycated hemoglobin (A1c) \citep{zaccardi2018glucose}. 

Since $2010$, the American Diabetes Association (ADA) has included measurement of A1c levels to both diagnosis and diabetes control \cite{american20186}. A1c levels reflect underlying glucose levels over the preceding $2–3$ months, testing is convenient because blood samples can be obtained at any time of day, overnight fasting is not required, and A1c within‐patient reproducibility is superior to that of fasting plasma glucose and oral glucose tolerance tests (OGTTs)  \cite{selvin2007short}. However, recent articles have provided evidence for the need to go beyond A1c and use new measures for glycemic control \citep{beyond2018need,Bergenstal1615} in order to capture more diverse aspects of the temporally evolving glucose levels beyond the average, for example, glucose variability and time in range metrics.  The time in metric range measures the proportion of time an individual's glucose levels is maintained in different target zones. In the case of diabetes, these can include ranges corresponding to hypoglycemia and hyperglycemia. In an innovative article, \cite{Beck400} validated the time in range metric, showing that it is a good predictor of long-term microvascular complications despite just measuring glucose values seven times per day.  \cite{Lu2370} reached similar conclusions but using CGM technology only for $24$ hours in each patient.  At the same time, it is well-known that two patients may have the same glycosylated hemoglobin and a completely different glycemic profile \citep{Beck994}. These new approaches and findings have lead clinical specialists to consider that continuous glucose measurement during long monitoring periods can lead to more accurate results in research and clinical practice than in standard methods \citep{Hirsch345}. In fact, since $2012$, the European Medicine Agency \citep{committee2012guideline} recommends the use of CGM to validate the effect of drugs for treatment or prevention of diabetes mellitus.  

Traditionally, CGM was designed for the risk management in real-time for type $1$ diabetes  and the control of glucose values with insulin pumps \citep{doi:10.1177/193229680900300106,FEIG20172347,DIMEGLIO20182449}. Notwithstanding, more recent applications of CGM have been more general. They involve, for example, screening patients, optimizing diet, epidemiological studies, assessing patient prognosis, and supporting treatment prescription, and have even been used in healthy populations \citep{Freeman134,10.1371/journal.pbio.2005143}. In addition to the increasing utility of CGM data, the technology is gradually becoming cheaper, and new devices capable of measuring glucose in a non-invasive way, for example, with glasses \citep{Nichols2013}, are quickly emerging. All of these advances are facilitating the adoption of CGM in standard clinical practice. 

In $2012$, a panel of experts discussed how to represent CGM data in an  ``easy to view format" \citep{doi:10.1089/dia.2013.0051}.  They also analyzed the convenience of using glycemic variability measures and other summary measures such as time in range to extract the recorded information from  CGM. In $2019$, ADA established an updated version of clinical standards to use and define target zones with time in range metrics \citep{Battelinodci190028}.  In a more recent review about the CGM metric, they establish time in range as a gold standard measure \citep{metricas}.  

\begin{figure}[ht!]
	
	\centering
	\includegraphics[width=0.7\linewidth]{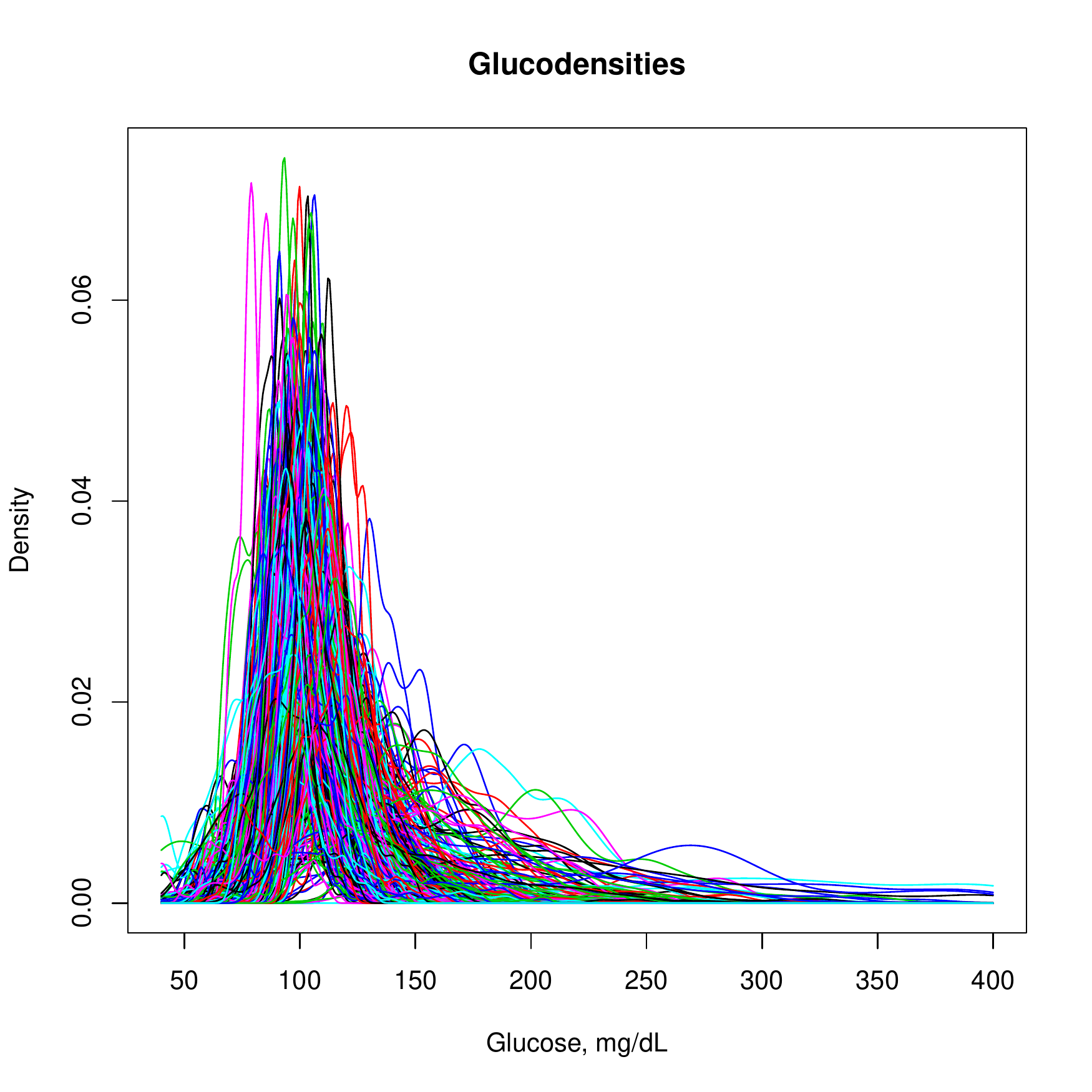}
	\caption{Example of a set of glucodensities estimated from a random sample of  the AEGIS population-based study} 
	\label{fig:graficodensidad}
\end{figure}

Motivated by the problem of analyzing data gathered via CGM more precisely, while still leveraging the advantages possessed by time in range metrics, we propose an approach based on the construction of a functional profile of glucose values for each subject.  Conceptually, the approach is a natural extension of time in range metrics in which the ranges shrink and size and increase in number, so that new profile effectively measures the proportion of time each patient spends at each specific glucose concentration rather than a coarsely defined range.  As a result of this, the new functional profile, which we refer to as a glucodensity, automatically and simultaneously captures all parameters arising from individual glucose distributions.   Figure \ref{fig:graficodensidad} illustrates a set of constructed glucodensities that represent the data objects for which we will propose the use of a tailored set of statistical methods. 

Mathematically, glucodensities constitute functional distributional data since each glucodensity represents a distribution of glucose concentrations.  As such, these complex and constrained curves cannot be directly analyzed with the usual techniques. To overcome this, we introduce a framework for the analysis of glucodensities by compiling suitable methods that are based on the calculation of glucodensities distances. We also reveal the superior clinical capacity of our representation compared to classical measures of diabetes.  
Finally, we demonstrate that our representation has a higher sensitivity than the standard time in range metric to explain the glycemic differences between patients in various settings, including regression analysis. A new shiny interface to use the methods outlined in this paper is available at \url{https://tec.citius.usc.es/diabetes}.

\begin{table}
			\centering
			\begin{tabular}{|c|c|}
						\hline 
						Biomarker & Clinical significance  \\ 
						\hline 
						A1c & \Centerstack{   Gold standard marker \\ in diabetes diagnosis and control}   \\
						\hline 
						HOMA-IR &  
						
						\Centerstack{	Measurements to  quantify insulin \\ resistance and $\beta$-cell function}

						\\ 
						\hline 
						CONGA &  \\   
						MODD &  \Centerstack{Summary indices of \\ glucose variability} \\   
						MAGE &    \\ 
						\hline 
					\end{tabular}
			\label{table:tablainicial}
			\caption{Clinical importance of biomarkers used in the statistical analysis}	\end{table}

\subsection{Outline}

The structure of this paper is as follows. First, we briefly describe the AEGIS study, and the methods used. We then formally introduce the concept of glucodensity, the estimation methods, and some essential statistical background to understand the statistical procedures introduced in the paper. Subsequently, we explain the regression models used in the validation of the representation. Afterward, we show the results that demonstrate the superiority of glucodensity over glucose representations of state-art. Then, we illustrate the use with real data of the glucodensities methodology in two-sample testing and cluster analysis.  Finally, we discuss the clinical implications of these results, their limitations, and the new perspectives of the glucodensities method in medicine and device technology.

	\section{Sample and Procedures}
\label{sec:material}
\subsubsection{Study design}

A subset of the subjects in the A Estrada Glycation and Inflammation Study (AEGIS; trial $NCT01796184$ at \url{www.clinicaltrials.gov}) provided the sample for the present work. In the latter cross-sectional study, an age-stratified random sample of the population (aged $\geq 18)$ was drawn from Spain's National Health System Registry. A detailed description has been published elsewhere \citep{Gude2017}. For a one year beginning in March, subjects were periodically examined at their primary care centre where they 
 $(i)$ completed an interviewer-administered structured questionnaire; $(ii)$ provided a lifestyle description; $(iii)$ were subjected to biochemical measurements; and $(iv)$ were prepared for CGM (lasting $6$ days). The subjects who made up the present sample were the $581$ ($361$ women, $220$ men) who completed at least $2$ days of monitoring, out of an original $622$  persons who consented to undergo a $6$-day period of CGM.  Another $41$ original subjects were withdrawn from the study due to non-compliance with protocol demands (n = $4$) or difficulties in handling the device (n = $37$). The characteristics of the participants are shown on the Table  \ref{table:tabla4}.

\begin{table}[ht]
	\centering
	\begin{tabular}{lll}
		\hline
		& Men $(n=220)$ & Women $(n=361)$ \\ 
		\hline
		Age, years & $47.8\pm 14.8$ & $48.2\pm14.5$ \\ 
		A1c, \% & $5.6\pm0.9$ & $5.5\pm0.7$ \\ 
		FPG $mg/dL$ & $97\pm23$ & $91\pm21$ \\ 
		HOMA-IR $mg/dL.\mu UI/m$ & $3.97\pm5.56$ & $2.74\pm2.47$ \\ 
		BMI $kg/m^2$ & $28.9\pm4.7$ & $27.7\pm5.3$ \\ 
		CONGA $mg/dL$  & $0.88\pm0.40$ & $0.86\pm0.36$ \\ 
		MAGE $mg/dL$ & $33.6\pm22.3$ & $31.2 \pm14.6$ \\ 
		MODD & $0.84\pm0.58$ & $0.77\pm0.33$ \\ 
		\hline
	\end{tabular} 
\caption{Characteristics of AEGIS study participants by sex. Mean and standard deviation are shown.	$BMI$ - body mass index; $FPG$ - fasting plasma glucose; $A1c$ - glycated haemoglobin; $HOMA-IR$ - homeostasis model assessment-insulin resistance; $CONGA$ - glycemic variability in terms of continuous overall net glycemic action; $MODD$ - mean of daily differences; $MAGE$ - mean amplitude of glycemic excursions.}
\label{table:tabla4}
\end{table}

\subsubsection{Ethical approval and informed consent}

The present study was reviewed and approved by the Clinical Research Ethics Committee from Galicia, Spain (CEIC$2012$-$025$). Written informed consent was obtained from each participant in the study, which conformed to the current Helsinki Declaration.

\subsubsection{Laboratory determinations}

Glucose was determined in plasma samples from fasting participants by the glucose oxidase peroxidase method. A1c was determined by high-performance liquid chromatography in a Menarini Diagnostics HA-$8160$ analyser; all A1c values were converted to DCCT-aligned values \citep{Hoelzel166}. Insulin resistance was estimated using the homeostasis model assessment method (HOMA-IR) as the fasting concentration of plasma insulin ($\mu$ units/mL) $\times$  plasma glucose (mg/dL)/ $405$  \citep{matthews1985homeostasis}.

\subsubsection{Glycaemic variability}

Glycaemic variability was measured in terms of continuous overall net glycemic action (CONGA) \citep{doi:10.1089/dia.2005.7.253}, the mean amplitude of glycaemic excursions (MAGE) \citep{MAGE}, and the mean of the daily differences (MODD) \citep{molnar1972day} in glucose concentration.

\subsubsection{CGM Procedures}

At the start of each monitoring period, a research nurse inserted a sensor (Enlite™, Medtronic, Inc, Northridge, CA, USA) subcutaneously into the subject's abdomen, and instructed him/her in the use of the iPro™ CGM device (Medtronic, Inc, Northridge, CA, USA). The sensor continuously measures the interstitial glucose level $40-400$ (range  mg/dL) of the subcutaneous tissue, recording values every $5$ min. Participants were also provided with a conventional OneTouchR VerioR Pro glucometer (LifeScan, Milpitas, CA, USA) as well as compatible lancets and test strips for calibrating the CGM. All subjects were asked to make at least three capillary blood glucose measurements (usually before main meals). These readings were taken without checking the current CGM reading. On the seventh day the sensor was removed and the data downloaded and stored for further analysis. If the number of data-acquisition ``skips" per day totalled more than $2$ h, the entire day's data were discarded.

\subsection{Time-in-range metric}

The time in the range metric was calculated with two different methods.  In the first, through the CGM records of the AEGIS study, we estimate the deciles of CGM records with normoglycemic patients and use as cut-offs the deciles  (Table \ref{table:cortes1}).  In the second, we use  cut-off points established by the ADA in the $2019$ Medical guideline \citep{Battelinodci190028} (Table \ref{table:cortes2}).

\begin{table}[ht!]

	\begin{center}

		\begin{tabular}{|c|c|}
			\hline 
			Range $1$	& $<85$ \\ 
			\hline 
			Range $2$	&  $85-90$ \\ 
			\hline 
			Range $3$	& $91-94$ \\ 
			\hline 
			Range $4$	&  $95-98$ \\ 
			\hline 
			Range $5$	& $99-101$ \\ 
						\hline 
						Range $6$	& $102-105$ \\ 
									\hline 
									Range $7$	& $106-109$ \\ 
												\hline 
												Range $8$	& $110-115$ \\ 
															\hline 
															Range $9$ &
															 $116-124$					\\ 
															 \hline 
															 Range $10$ &
															 $>125$				
\\
			\hline
		\end{tabular} 
	\end{center}
\caption{Cut-offs  for metric time in range  using own estimations throught  normoglucemic individuals of AEGIS study}
 \label{table:cortes1}
\end{table}

\begin{table}[ht!]

	\begin{center}

		\begin{tabular}{|c|c|}
			\hline 
			Range $1$	& $<54$ \\ 
			\hline 
			Range $2$	&  $54-69$ \\ 
			\hline 
			Range $3$	& $70-180$ \\ 
			\hline 
			Range $4$	&  $181-250$ \\ 
			\hline 
			Range $5$	& $>250$ \\ 
			\hline 
		\end{tabular} 
	\end{center}
	\caption{Cut-offs  for metric time in range  following ADA guidelines
		\cite{Battelinodci190028}}
	\label{table:cortes2}
\end{table}

	\subsubsection{Statistical procedure}

The density functions for each individual was estimated with non-parametric Nadaraya-Watson procedure. For this purpose, we used a Gaussian kernel and rule of thumb as a smoothing parameter. In addition, we estimate quantile representation for $2$-Wasserstein methods using the empirical distribution.

The following three regression models were used: i) The non-parametric kernel functional regression model through $2$-Wasserstein distance with the glucodensity as predictor \citep{ferraty2006nonparametric}; ii) A global $2$-Wasserstein regression model where the glucodensity is response \citep{petersen2019}; and iii)  $k$-nearest neighbor algorithm in the case of time in range metrics with $k=10$ neighbors.

In the case of time in range metrics, we applied the isometric log-ratio (ilr) transformation for compositional data prior to fitting the model. To avoid problems with zeros, a fixed positive constant was added to each each range, which were then normalized to add to $1$.

All analyses were carried out using R software. Functional data analysis was performed using the \texttt{fda.usc} package \citep{Febrero-Bande2012}, which is freely available at  \url{https://cran.r-project.org/}, and our own implementations of the ANOVA test of \cite{dubey19} or Fr\`{e}chet regression in \cite{petersen2019} using the $2$-Wasserstein distance. The glucodensities and their quantile representation were estimated using the R basis functions.

%

\section{Definition and Estimation of the Glucodensity}
\label{sec:methods}
For patient $i$, denote the gathered glucose monitoring data by pairs $(t_{ij}, X_{ij})$, $j = 1,\ldots,m_i,$ where the $t_{ij}$ represent recording times that are typically equally spaced across the observation interval, and $X_{ij}$ is the glucose level at time $t_{ij}\in [0,T_i].$ Note that the number of records $m_i$, the spacing between them, and the overall observation length $T_i$ can vary by patient.  One can think of these data as discrete observations of a continuous latent process $Y_i(t),$ with $X_{ij} = Y_i(t_{ij}).$  The glucodensity for this patient is defined in terms of this latent process as $f_i(x) = F_i'(x),$ where
$$
F_i(x) = \frac{1}{T_i}\int_0^{T_i} \mathbf{1}\left(Y_i(t) \leq x\right) \mathrm{d}t \quad \text{for } \, \quad \inf_{t \in [0,T_i]} Y_i(t) \leq x \leq \sup_{t \in [0,T_i]} Y_i(t)
$$
is the proportion of the observation interval in which the glucose levels remain below $x.$ Since $F_i$ are increasing from 0 to 1, the data to be modeled are a set of probability density functions $f_i,$ $i = 1,\ldots,n.$  

Of course, neither $F_i$ nor the glucodensity $f_i$ is observed in practice, but one can construct an approximation through a density estimate $\tilde{f}_{i}(\cdot)$ obtained from the observed sample. In this case of CGM data, the glucodensities may have different support and shape. Therefore, we suggest using a non-parametric approach to estimate each density function. For example, using a kernel-type estimator, we have  
\begin{equation*}
	\tilde{f}_{i}(x)= \frac{1}{m_i} \sum_{j=1}^{m_i} K_{h_i}(x-X_{ij}),
\end{equation*}
where $h_{i}>0$ is the smoothing parameter and $K_{h_i}(s)= \frac{1}{h_i}K(\frac{s}{h_i})$. The choice of $K$ does not have a big impact on the efficiency of the estimator, but the value of $h_i$ is crucial.

Several alternatives for selecting the smoothing parameter have been proposed in the literature, including cross-validation, minimizing the estimated mean integrated squared error (MISE), or a ``rule of thumb" derived from the assumption that the density is Gaussian.  In this last case, the choice can be explicitly written as $\tilde{h}_i= 1.06 \tilde{\sigma_{i}}m_i^{-1/5}$, where $\tilde{\sigma_{i}}$ is the sample standard deviation of the $X_{ij}$. For more details, see \cite{Silverman86}. Other approaches for the density function estimation include the use orthogonal series (e.g., Fourier or Wavelet) expansions, splines, or smoothing of histograms. For further details the reader is referred to \cite{Antoniadis1997,doi:10.1080/01621459.1991.10475021,muller2014density}.

\subsection{Distance-based Descriptive Statistics}

Let $[a,b]$ be an interval of the real line, which may be unbounded, and suppose that each glucodensity $f_i$ has support contained in $[a,b]$. From a statistical point of view, the sample $f_1,\ldots,f_n$ may be modeled and analyzed using methods of functional data analysis \citep{ramsay2005functional,wang2016functional}.  However, since the $f_i$ must be positive and satisfy $\int_a^b f_i(x) dx = 1,$ classical methods have in recent years been adapted to account for the nonlinear, distributional structure of density samples \citep{petersen2016,hron2016simplicial}.  The general approach is to define a metric or distance between densities that, in turn, leads to descriptive statistics that respect the unique density properties.  For example, define the data space of glucodensities as $A:= \{f:[a,b]\to \mathbb{R}^{+}: \int_{a}^{b} f(x)dx=1 \hspace{0.2cm} \text{and} \hspace{0.2cm} \int_{a}^{b} x^2f(x)dx < \infty \}$. Given  two  arbitrary  glucodensities $f,g\in A$, the $2$-Wasserstein distance \citep{villani2008optimal} between $f$ and $g$ is
\begin{equation}
\label{eq: Wdis}
d_{W^{2}}(f,g)=  \sqrt{\int_{a}^{b} (F^{-1}(x)-G^{-1}(x))^{2}}dx,
\end{equation}
where $F$ and $G$ are the cumulative distribution functions (cdfs) of the density functions $f$ and $g$.

The $2$-Wasserstein distance is a natural distance to measure the similarity between density functions through its representation in the space of the quantile (inverse cdf) functions and it has already been successfully applied in biological problems. Furthermore, it has computational and modeling advantages compared to the usual $L^2[a,b]$ metric when glucodensities have different support within $[a,b]$. Finally, it has a physical interpretation in the theory of optimal transport.

As glucodensities are distributional data, the subsequent application of the usual techniques for functional data, such as estimation of mean, covariance, and regression models, may lead to misleading results. Hence, we have chosen to use models based on the $2$-Wasserstein distance, although other choices are possible.
As a starting point, based on the notion of distance we can generalise the mean and variance of a random variable that takes values in an abstract space with metric structure \citep{AIHP_1948__10_4_215_0}.  As we will see, similar adaptations can be developed for regression, hypothesis testing, or to perform cluster analysis. Given a distance $d: A\times A\to \mathbb{R}^{+},$ of which $d_{W^2}$ is one example, and a random variable $f$ defined on $A$, the \emph{Fr\'echet mean} of $f$ is
 \begin{equation*}
\mu_{f}= \argmin_{g\in A} E(d^{2}(f,g)).
 \end{equation*}
The \emph{Fr\'echet variance} of $Z$ is then
 \begin{equation*}
 \sigma_f^{2}=  E(d^{2}(f,\mu_{f})).
 \end{equation*}
If the choice of distance is the Wasserstein metric $d_{W^2},$ these are given the names of Wasserstein mean and variance, respectively.  In the following subsections we will extend these concepts of Fr\`echet to statistical methodologies of regression, clustering, and hypothesis testing based on the notion of distance.

\section{Regression models with glucodensities}

\subsubsection{Non-parametric regression model with Glucodensity as the Predictor}

Let $f$ be a functional random variable taking values in $(A,d_{W^{2}})$  and $Y$ a random  variable that take values in the real line. We assume the following regression relationship between $f$ and $Y$, which represent the predictor and response variables, respectively:
\begin{equation}
Y= g(f)+\epsilon
\end{equation}
where $g:A\to \mathbb{R}$ is an unknown  smooth function, and the random error $\epsilon$ satisfies $E(\epsilon)=0$.

Given a sample  $\{(f_i,Y_i)\in A\times \mathbb{R}\}_{i=1}^{n}$, most non-parametric estimators $\tilde{g}(\cdot)$ have the form of a weighted average of the responses
\begin{equation}
\tilde{g}(x)= \sum_{i=1}^{n} w_{ni}(x)Y_i.
\end{equation}
In general, the weights $w_{ni}(x)$ depend on the distance between each $f_i$ and $x$, with larger distances receiving lower weights, and satisfy $\sum_{i = 1}^n w_{ni}(x) = 1$  \citep{ferraty2006nonparametric}.  A typical choice would be the Nadaraya–Watson weights 
\begin{equation}
w_{ni}(x)= \frac{K(\frac{d(x,f_i)}{h})}{\sum_{i=1}^{n}(K(\frac{d(x,f_j)}{h}))},
\end{equation}
where 
$h$ is a smoothing parameter and $K: \mathbb{R} \to \mathbb{R}$ is a known univariate probability density function called the kernel. For more details about this procedure see \cite{ferraty2006nonparametric}.  As an alternative for the above method, we can use the kernel methods in Reproductive Kernel Hilbert Spaces (RKHS) \citep{PREDA2007829,10.5555/2946645.3053434}.

\subsection{Regression model with Glucodensity as the Response}

In the case of the regression models with a density function as response, the literature is not very extensive to the current date \citep{NERINI20074984,doi:10.1080/01621459.2019.1604365,petersen2019,capitaine2019frchet,talska2018compositional}.  In this article we use the model proposed in \cite{petersen2019} which allows us to incorporate the desired metric $d_{W^2}$ and is a direct generalization of classical linear regression.  The primary rationale for our use of this model is that, unlike the other approaches mentioned above, there is a methodology developed to performance inferential procedures such as confidence bands and hypothesis testing in order to establish the significance of the input variables in the model \cite{alex2019wasserstein}.

Let $f$ be a random variable (e.g.\ a glucodensity) that take values in the space of  $(A,d_{W^2})$ defined above. Consider a random vector $U \subset \mathbb{R}^d$ that contains the set of predictors. Our interest is in the Fr\`echet regression function, or function of conditional Fr\`echet means,
\begin{equation}
\overline{f}(u):= \argmin_{g\in A} E(d^{2}_{W^2}(f,g)|U=u), \quad u\in \mathbb{R}^{d}
\end{equation}
\cite{petersen2019} imposes a particular model for $\overline{f}$ that, in direct analogy to classical linear regression, takes the form of a weighted Fr\`echet mean:
\begin{equation}
\label{eq: alexModel}
\overline{f}(u)= \argmin_{g\in A} E(s(U,u) d^{2}_{W^2}(f,g)),  u\in \mathbb{R}^{d}.
\end{equation}
Here, the weight function is 
\begin{equation}
s(U,u)= 1+(U-\mu)^{T}\Sigma^{-1}(u-\mu), \hspace{0.2cm} \mu= E(U), \Sigma= \mathrm{Cov}(U),
\end{equation} 
and $\Sigma$ is assumed to be positive definite.

Given a sample $(U_i, f_i),$ $i = 1,\ldots,n,$ of independent pairs each distributed as $(U, f),$ one can proceed to estimate $\overline{f}(u)$ for any desired input $u.$  Due to the intimate connection between the Wasserstein metric and quantile functions as in \eqref{eq: Wdis}, for most inferential procedures it is sufficient to estimate the conditional Wasserstein mean quantile function $\overline{Q}(u)$ corresponding to $\overline{f}(u).$  Let $D$ be the set of quantile functions, $Q_i$ the quantile function corresponding to the random density $f_i,$ and define empirical weights $s_{in}(u) = 1 + (U_i - \overline{U})^T\hat{\Sigma}^{-1}(u - \overline{U}),$ where $\overline{U}$ and $\tilde{\Sigma}$ are the sample mean and variance of the $U_i,$ respectively.  The natural estimator under $d_{W_2}$ is the weighted empirical mean quantile function  
\begin{equation}
\tilde{\overline {Q}}(u)= \argmin_{Q\in D} \sum_{i=1}^{n} s_{in}(x) \lVert Q-Q_i\rVert^{2},
\end{equation}
where $\lVert\cdot\rVert$ denotes the $L^{2}[0,1]$ norm on $D$.

A straightforward algorithm for computing  $\tilde{\overline {Q}}(u)$ is shown in Supplementary Material of original reference \cite{alex2019wasserstein}. In addition, two algorithms are given to estimate the confidence bands at a given significance level $\alpha$ for both the quantile functional parameter $\overline {Q}(u)$ and the density parameter $\overline {f}(u)$.

\section{Results}

\subsection{Clinical validation of the glucodensity}

To validate the glucodensity representation, we use the database from the AEGIS study \citep{Gude2017}. The database contains the continuous glucose monitoring data between $2$-$6$ days of $581$ patients from a random sample of a general population.  A detailed description of the data is introduced in Section \ref{sec:material} together with characteristics of patients in Table \ref{table:tabla4}. To develop the validation task, we use two different regression models: i) a non-parametric regression model where the unique predictor is glucodensity, and ii) a linear regression model where the response is a glucodensity. Further details on the regression models used can be found in the  Section \ref{sec:methods}. The first model was used to predict glycated hemoglobin  (A1c) \citep{Kilpatrick335}, homeostatic model assessment (HOMA-IR) \cite{Ausk1179}, and the following measures of glycemic variability
\cite{Service1398,variabilidadmon,Gude2017}: continuous overall net glycemic action (CONGA), mean amplitude of glycemic excursions (MAGE) and mean of daily differences (MODD), through glucodensity representation. In contrast, the second was used to predict the glucodensity with the five variables above.  Figure~\ref{fig:graficodensidad} gives a visualization of the sample of glucodensities used in these models. Biological significance in variables under consideration  is described in Table \ref{table:tablainicial}.

\subsubsection{Prediction of biomarkers using the glucodensity}

\begin{figure}[ht!]
	\centering
	\includegraphics[width=0.7\linewidth]{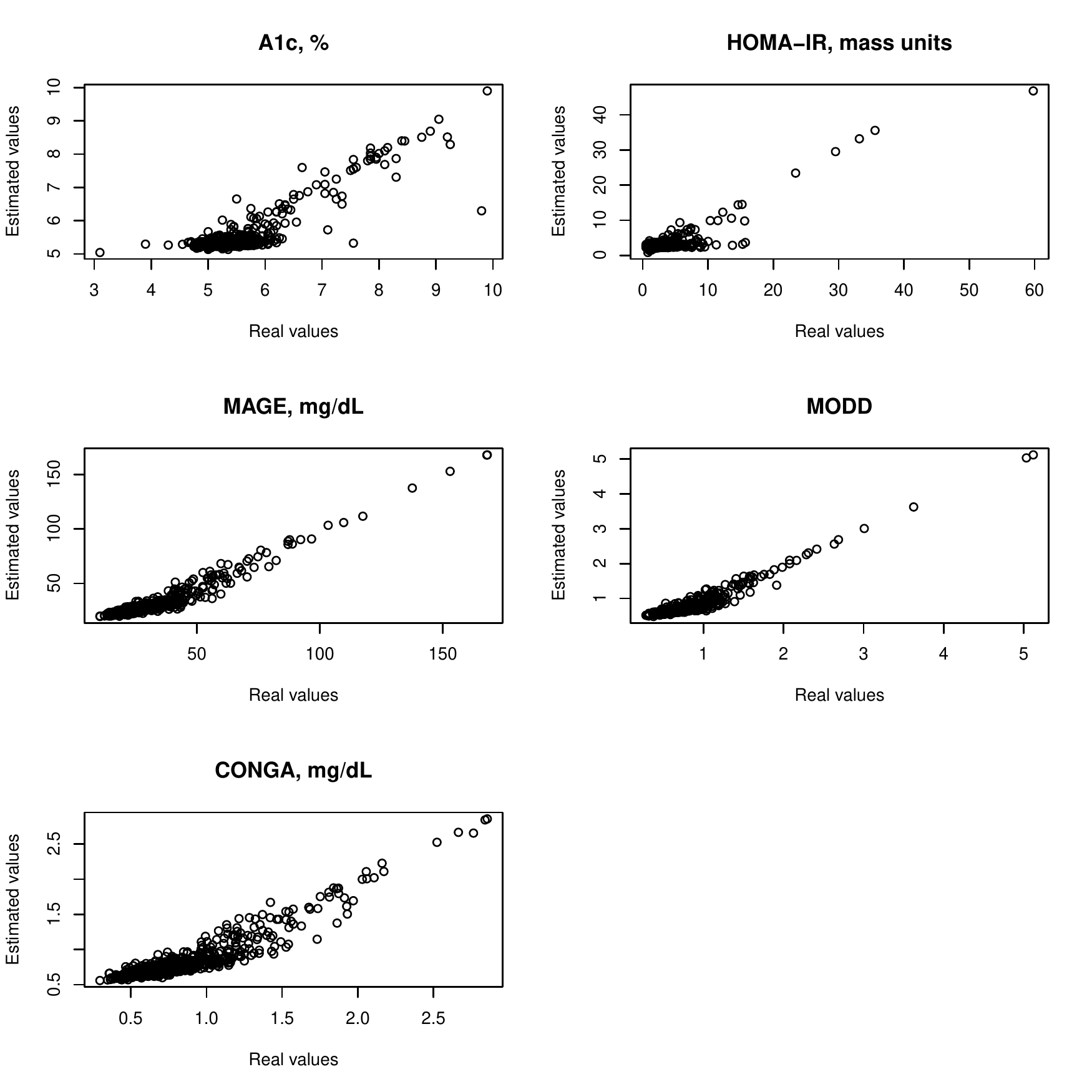}
	\caption{Real values vs Estimated values when glucodensity is predictor}
	\label{fig:reg1}
\end{figure}

The aim of the first set of regression analyses is to demonstrate that the glucodensity is sufficiently rich in its information content to recover the aforementioned biomarkers with high precision. To quantify this precision, we estimated the $R^2$ after fitting a non-parametric model for each biomarker as the outcome variable, using the glucodensity as the sole predictor (i.e. independent variable). The $R^2$ estimates for A1c, HOMA-IR, MAGE, MODD, CONGA  were $0.79$, $0.79$, $0.92$, $0.86$, and $0.92$ respectively. To supplement the results, Figure \ref{fig:reg1} shows the predicted values against the observed values, where the outstanding predictive capacity of the glucodensity can be seen independently of high or low response values.

\subsubsection{Prediction of the glucodensity using biomarkers}

\begin{figure}[ht!]
	\centering
	\includegraphics[width=0.7\linewidth]{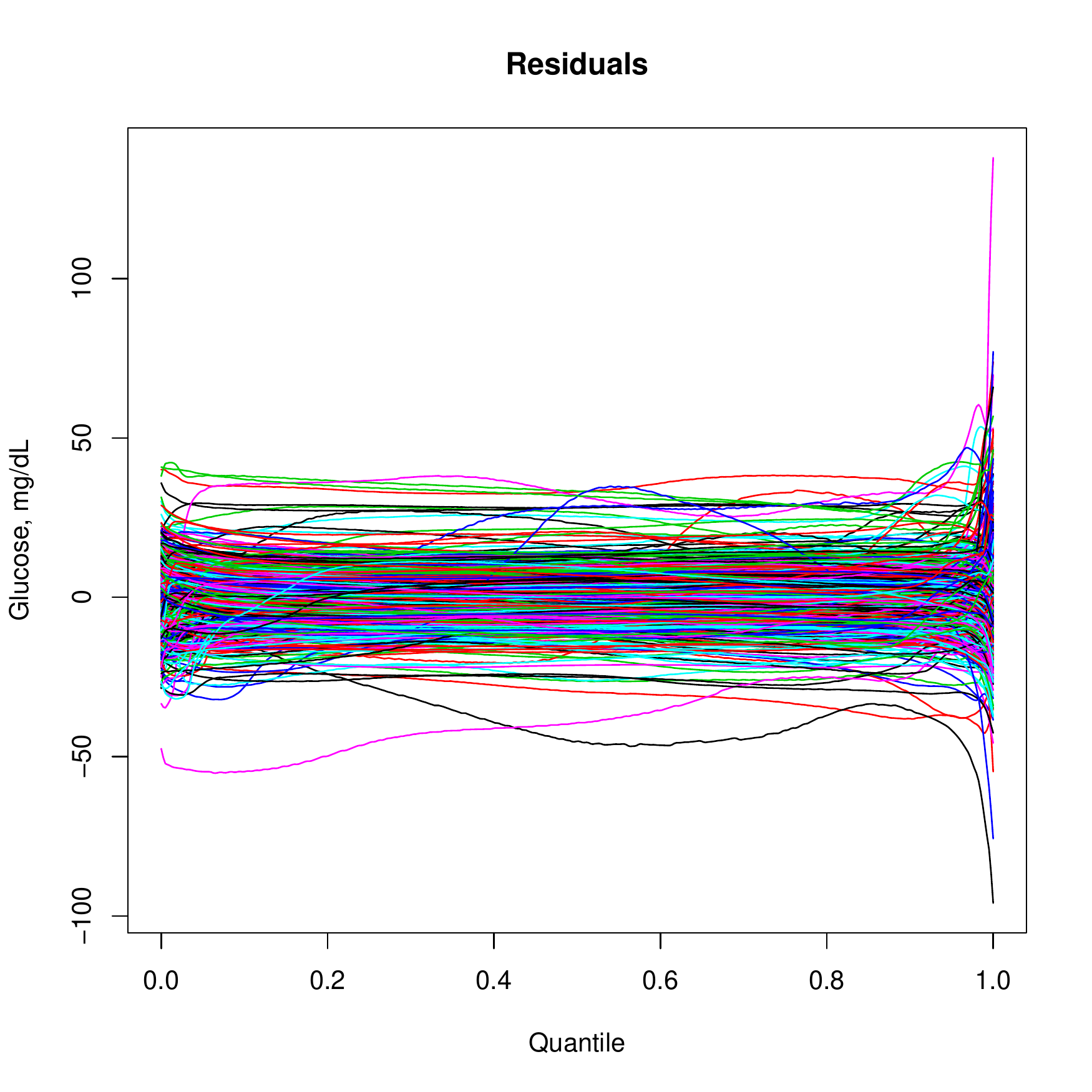}
	\caption{Residuals in quantile space prediction glucodensities}
	\label{fig:reg2}
\end{figure}

In the second regression analysis with the glucodensity as the outcome variable, we aim to show that the previous measurements commonly used in the clinical practice are not capable of capturing the glucodensity with high accuracy. This fact is not completely
surprising because, as noted by some authors \citep{zaccardi2018glucose}, the information provided by a CGM is more precise than that contained in summary measures. To accomplish this, we computed a suitable version of $R^{2}$ for this task after fitting a regression model where the response is a glucodensity, and the previous variables are the predictors. In this case, the $R^2$ estimated was $0.74$. As predicted, compared to the previous section's results, we were not able to accurately capture the complex nature of glucodensities, even while using the combined predictive power of several commonly used summary measures. Moreover, in some cases, the differences in prediction can be significant (see Figure \ref{fig:reg2}).     

\subsection{Comparison of time in range metrics with glucodensities}

To illustrate the higher clinical sensitivity of glucodensities compared to time in range metrics, we compared the ability of each representation to predict A1c, HOMA-IR, and glycemic variability metrics MODD, MAGE, and CONGA, using the data from AEGIS study. The predictive capacity of the glucodensity representation was illustrated above, and this section gives the corresponding results for time in range metrics, where these were calculated according to two sets of cutoffs. In the first, the deciles of the normoglycemic individuals from the AEGIS study were used, while in the second those proposed by the ADA were used.  Tables \ref{table:cortes1} and \ref{table:cortes2} in Section \ref{sec:material} show the exact cutoff values for both cases. Since the time in range metrics constitute a sample of standard compositional data, the isometric log-ratio (ilr) transformation was employed in combination with a $k$-nearest neighbor algorithm as a regression model for predicting the scalar variables. Methodological details about this statistical procedure can be found in Section \ref{sec:material}.

\subsubsection{Prediction of A1c, HOMA-IR and glycemic variability measures using time in range metrics}

\begin{figure}[ht!]
	\centering
	\includegraphics[width=0.7\linewidth]{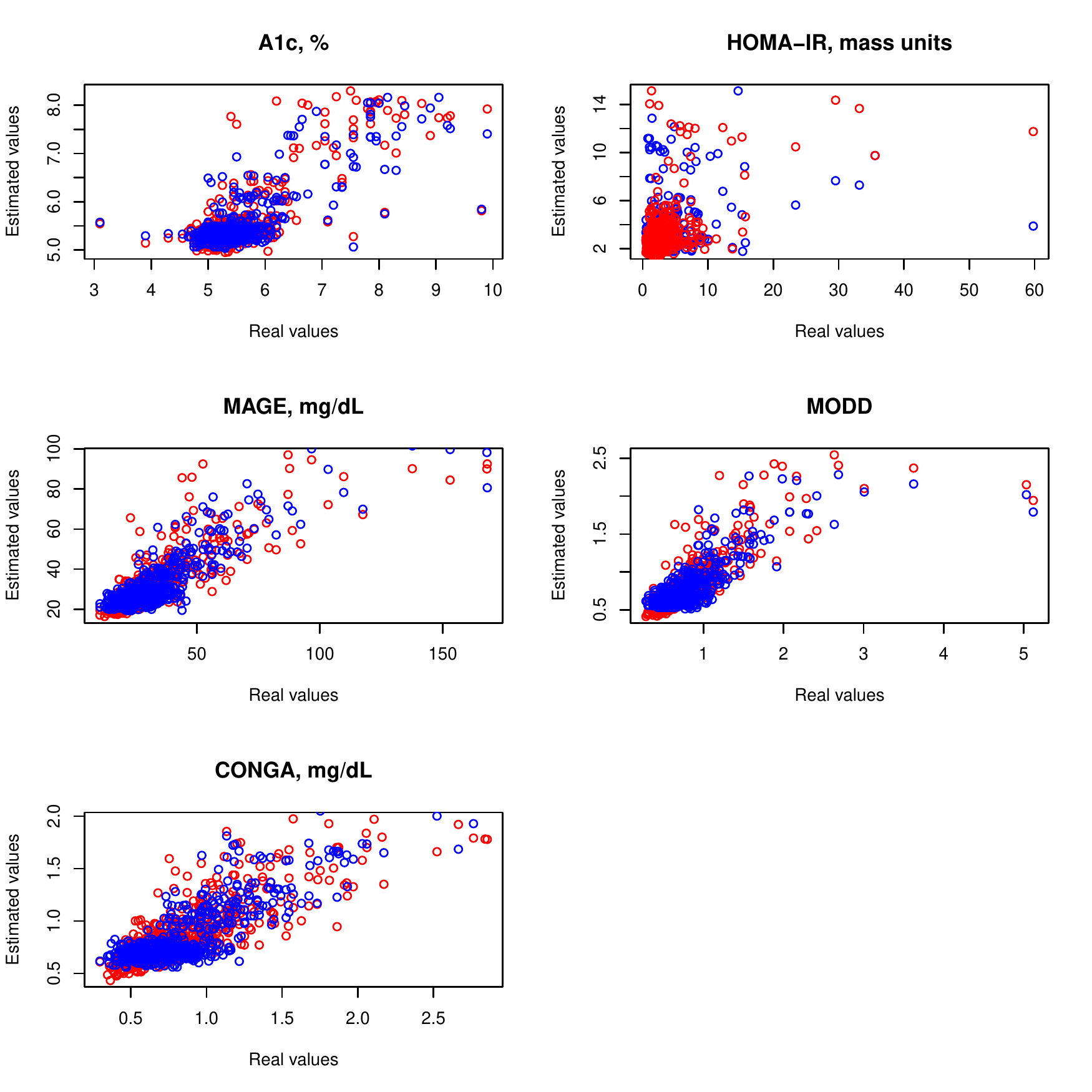}
	\caption{Real values vs. Estimated values when time in range metric is the predictor.  Blue, time in range metric with cut-offs calculated with normoglycemic patients of AEGIS database. Red, time in range metric using of cut-offs suggested by ADA.}
	\label{fig:graficas2}
\end{figure}

Figure \ref{fig:graficas2} compares the real and estimated values of the previous five variables under the two time in range metrics under consideration with.   
Table \ref{table:resultspuntos} provides the estimates of $R^{2}$ for each variable and metric.  
\begin{table}
	\begin{center}

		\begin{tabular}{|c|c|c|c|c|c|}
			\hline 
			& A1c    & HOMA-IR  & CONGA  & MAGE  & MODD \\ \hline 
			Normoglycemic  cut-off & $0.63$     & $0.22$ & $0.68$  & $0.65$ & $0.65$ \\ 
			\hline 
			ADA cut-off         
			& $0.61$  & $0.08$  & $0.73 $  & $0.69$  & $0.60$  \\ 
			\hline 
		\end{tabular}
	\end{center} 
	\caption{$R^{2}$ estimated with time in range metrics under consideration}
	\label{table:resultspuntos}
\end{table}
The predictive capacity is significantly worse than that attained by the glucodensity methodology.  The superiority of the glucodensity is particularly noteworthy in the case of the HOMA-IR variable, where the association is quite weak for time in range metrics. Even for the other variables where the values of $R^2$ are moderate, the larger residuals seen in diabetes patients with more severe alterations of glucose metabolism indicate that time in range metrics are particularly poorly suited for such patients.  Interestingly, we do not observe substantial or consistent differences between the two time in range metrics used, as deciles perform better than ADA criteria for two of the variables, while in other instances the ordering was reversed.

\section{Hyphotesis and clustering analysis with glucodensities}

\subsection{Analysis of variance with glucodensities}
As a special case of regression, suppose we have a sample $f_1$, $\ldots$ $f_n$ of glucodensities defined on  $(A,d_{W})$ belonging to $k$ different groups  $G_1,G_2,\cdots,G_k$ that partition $\{1,\ldots,n\}$ and are of size $n_j$ $(j=1,\cdots,k)$, so that $\sum_{j=1}^{k} n_j=n$. If the goal is to simply test whether the Wasserstein means are equal for each group, \cite{alex2019wasserstein} developed testing procedures based on model \eqref{eq: alexModel} for this purpose. An advantage of this model is its flexibility, which allows for multiple factor layouts as well as tests for interactions.  However, the theoretical properties of these tests require a type of equal variance assumption that may be restrictive for some data sets.

More generally, one may wish to test the null hypothesis that the population distributions of the $k$ groups share common Wasserstein means and variances, against the alternative that at least one of the groups has a different population distribution compared to the others in terms of either its Wasserstein mean or variance. In this scenario, \cite{dubey19} investigated a test statistic
based on the group proportions $\lambda_{j,n} = n_j n^{-1}$, the groupwise sample Wasserstein means $\tilde{\mu}_j= \argmin_{g \in A} \sum_{i\in G_j} d_{W^2}^2(f_i,g)$ and variances $\tilde{V}_j = n_j^{-1}\sum_{i \in G_j} d_{W^2}^2(f_i, \tilde{\mu}_j),$ the pooled Wasserstein mean $\hat{\mu}_p = \argmin_{g \in A} \sum_{j = 1}^k \sum_{i \in G_j} d^2_{W^2}(f_i, g)$ and variance $\tilde{V}_p = n^{-1}\sum_{j = 1}^k\sum_{i \in G_j} d^2_{W^2}(f_i, \tilde{\mu}_p),$ and finally the quantities
$$
\tilde{\sigma}_{j}^2= \frac{1}{n_j}\sum_{i\in G_j}^{} d^{4}_{W^2}(f_i,\hat{\mu_j})-\left\{\frac{1}{n_j}\sum_{i\in G_j} d^{2}_{W^2}(f_i,\hat{\mu_j})\right\}^{2}
$$
as estimates of the variance of $\hat{V}_j.$
Then, with
\begin{equation*}
F_n= \tilde{V}_p-\sum_{j=1}^{k} \lambda_{j,n} \tilde{V}_j, \quad
R_n = \sum_{j<l}^{} \frac{\lambda_{j,n},\lambda_{l,n}}{\tilde{\sigma}^{2}_l \tilde{\sigma}^{2}_j}(\tilde{V}_j-\tilde{V}_l),
\end{equation*}
the proposed test statistic is
\begin{equation}
T_n= \frac{nR_n}{\sum_{j=1}^{k}\lambda_{j,n}\tilde{\sigma}^{-2}_j}+\frac{nF^2_n}{\sum_{j=1}^{k}\lambda_{j,n}\tilde{\sigma}^{2}_j}.
\end{equation}

\cite{dubey19} demonstrated that the corresponding test is distribution-free, in that the limiting distribution of $T_n$ does not depend on the underlying distribution under some assumptions.  In practice, it was also demonstrated that it can be useful to calibrate the test under the null hypothesis via a simple empirical bootstrap over the preceding statistics. For more details we refer the reader to the supplementary material of the original reference.

\subsection{Energy distance methods with glucodensities}

The energy distance is a statistical distance between two distribution functions proposed in  1984 by Gábor J. Székely (\cite{szekely2017energy}). This distance is inspired by the concept of gravitational energy between two bodies, and has experienced a rise in appeal for modern statistical applications due to its applicability to data of a complex nature such as functions, graphs or objects that live in negative type space.

Consider independent random variables $Y,Y^{\prime}\sim F$ and $Z,Z^{\prime}\sim G$ that are defined on a (semi)metric space $(\Omega, \rho)$ of negative type, where $\rho: V\times V\to \mathbb{R}$ is the semi-metric.  Though the notation in this section is quite general, in particular we have in mind the case $(\Omega, \rho) = (A, d_{W^2})$ corresponding to glucodensities.  The energy distance associated with $\rho$ between the distribution $F$ and $G$ is
\begin{equation*}
\epsilon_{\rho}(F,G)=  2 E(\rho(Y,Z))-E(\rho(Y,Y^{\prime}))-E(\rho(Z,Z^{\prime})).
\end{equation*}   

Given random samples $Y_1,\dots,Y_n \overset{\mathrm{iid}}{\sim} F$ and $Z_1,\dots,Z_m \overset{\mathrm{iid}}{\sim} G$, the sample energy distance is
\begin{equation*}
\tilde{\epsilon}_{\rho}(F,G)=  2 \frac{1}{nm}  \sum_{i=1}^{n} \sum_{j=1}^{m} \rho(Y_i,Z_j)- \frac{1}{n^{2}} \sum_{i=1}^{n} \sum_{i=1}^{n} \rho(Y_i,Y_j) - \frac{1}{m^{2}} \sum_{i=1}^{m} \sum_{i=1}^{m} \rho(Z_i,Z_j)   .
\end{equation*}

The asymptotic distribution of the above statistic for a null hypothesis $(H_0:F=G)$ as well as for the alternative $(H_a:F\neq G)$ is dependent on the chosen semimetric $\rho$. Besides, its expression is difficult to calculate and to implement in practice. Hence, when using the energy distance based methods, the distribution under the null hypothesis is usually calibrated with a permutation method.  Alternatives to calibrate the distribution under the null hypothesis include the wild or a weighted boostrap, as described in \cite{LEUCHT2013257,JIMENEZGAMERO2019131}.  The energy distance can also be extended to handle samples from more than two populations.  Given $k$ independent samples $Y_{j1},\ldots, Y_{jn_j}\overset{\mathrm{iid}}{\sim} F_j$, $j = 1,\ldots,k,$ the energy distance statistic is
\begin{equation}
\tilde{\epsilon}_{\rho}(F_1,\ldots, F_k) \sum_{1\leq j<l\leq k}^{} \frac{n_j n_l}{2n}  [2g_{jl}-g_{jj}-g_{ll}], \,\, g_{jl} = \frac{1}{n_jn_l} \sum_{i = 1}^{n_j}\sum_{i' = 1}^{n_l} \rho(Y_{ji}, Y_{li'}),
\end{equation}
where $n= n_1+\cdots, n_k.$

We now explain how this statistic can be adapted to perform clustering.  
Consider random pairs $(Y_i, I_i),$ $i = 1,\ldots,n,$ where $Y_i$ is observed and takes values in $(\Omega, \rho),$ while $I_i \in \{1,\ldots,k\}$ is an unobserved label of cluster membership.  The task is to recover the true clusters $C_j^* = \{i: I_i = j\},$ $j = 1,\ldots,k.$  Let $C_1,\ldots,C_k$ be a generic partition of $\{1,\ldots,n\}$, and denote the size of each cluster by $|C_j|$. Then a clustering may be chosen by optimizing the statistic 
\begin{equation}
S_\rho(C_1,\ldots, C_k) = \sum_{1\leq j<l\leq k}^{} \frac{n_j n_l}{2n}  [2\tilde{g}_{jl}-\tilde{g}_{jj}-\tilde{g}_{ll}], \,\, \tilde{g}_{jl} = \frac{1}{|C_j||C_l|}\sum_{(i,i') \in C_j\times C_l} \rho(Y_i, Y_i')
\end{equation}
over all possible clusters $C_j.$  At first view, this seems computationally intractable due to the appearance of distances between the elements of each cluster. However, defining
\begin{equation}
W_\rho(C_1,\ldots,C_k) = \sum_{j=1}^{k} \frac{|C_j|}{2} \tilde{g}_{jj}, \\
\end{equation}
it can be proven that $S_\rho + W_\rho$ is constant. This implies that maximizing $S_\rho$ is equivalent to minimizing $W_\rho$.

In \cite{9103121}, the authors show the equivalence between the previous optimization problem with the clustering procedure kernel $k$-means. The latter relationship allows the solving of kernel Kgroup clustering procedure through the popular heuristics algorithms as  Hartigan and Lloyd that allow finding the optimal solution with the k-means algorithm.

\subsection{ Example of hypothesis testing and clustering analysis with glucodensity  methodology}

Below, we illustrate the methodology of glucodensities in hypothesis testing and cluster analysis with the $2$-Wasserstein distance. We use the ANOVA test \citep{dubey19} and the $k$-groups algorithm \cite{9103121}.

\subsubsection{Hypothesis testing}

An interesting question to address in an epidemiological study is whether there are differences between men and women in the glycemic profile. The ANOVA test is an important instrument to establish whether there are statistically significant differences in mean and variance with glucodensities where there are two or more patient groups. After applying this method with AEGIS data, the test yields a p-value equal to $0.10$. Therefore, there is no statistically significant difference between men and women at the significance level of $5$ percent.

\begin{figure}[ht!]
	\centering
	\includegraphics[width=0.7\linewidth]{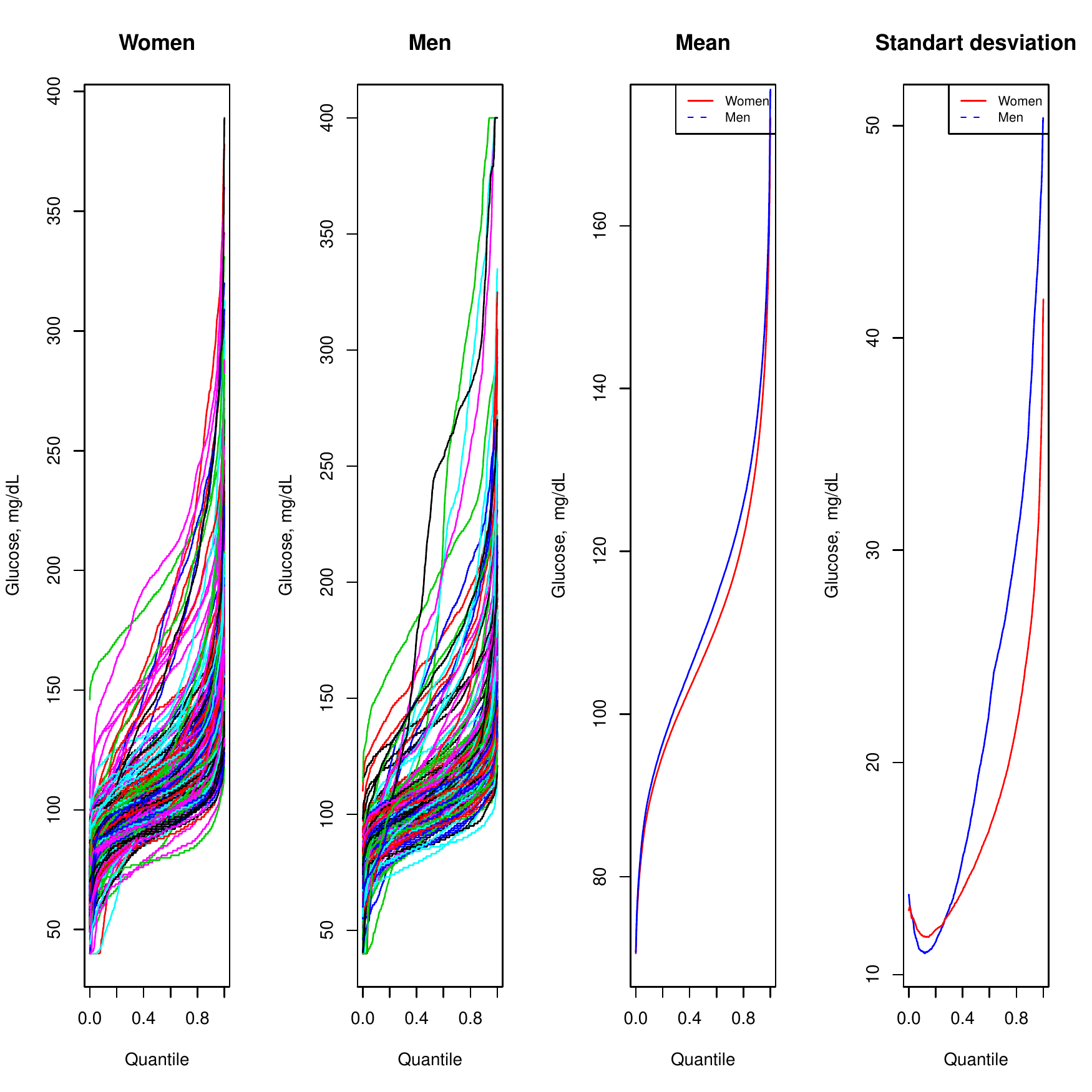}
	\caption{(Left two panels) Glucodensities for men and women of the AEGIS study, plotted as quantile functions; (Third panel) 2-Wasserstein mean quantile functions for each group; (Fourth Panel) Cross-sectional standard deviation curves for quantile functions in each group.}
	\label{fig:graficostest}
\end{figure}

Figure \ref{fig:graficostest} shows the glucodensity samples for each gender using their quantile representations.  The pointwise means of these quantile functions constitute the quantile function of the sample Wasserstein mean glucodensites.  These, together with pointwise standard deviation curves, are also shown in Figure~\ref{fig:graficostest}. On average, the groups are quite similar. However, certain discrepancies are observed between both groups in terms of their variance, although not large enough for the test to show statistical significant differences.

\subsubsection{Clustering analysis}

Cluster analysis is an essential tool for identifying subgroups of patients with similar characteristics. As an example, with the diabetes patients' data from the AEGIS study, we perform a cluster analysis using three clusters. To establish when a patient has diabetes, we use the doctor's previous diagnostic criteria, or if individuals currently have their glucose values measured with A1c and FPG in the ranges established by the ADA to be classified in that category. 

\begin{figure}[ht!]
	\centering
	\includegraphics[width=0.7\linewidth]{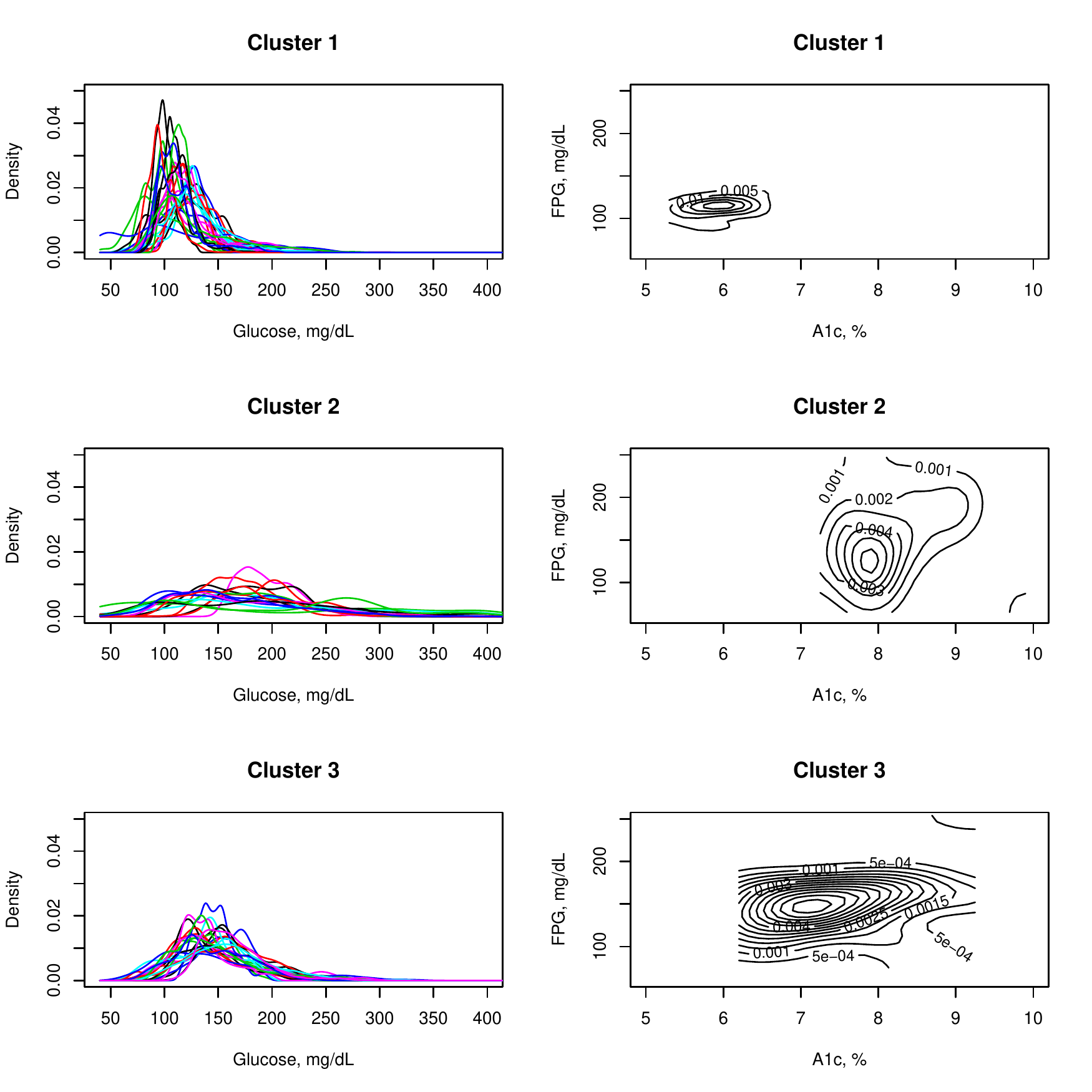}
	\caption{Clustering analysis of diabetes patients in AEGIS study}
	\label{fig:cluster}
\end{figure}

Figure \ref{fig:cluster} contains the results of applying the cluster analysis in diabetes patients. The algorithm has identified three differentiated groups of patients. The first group is patients with normal glucose values, probably because they are on medication, and the diagnosis of diabetes was made in the past. The second group are patients with slightly altered diabetes metabolism. Finally, the last group is patients with severely altered glucose values, and as can be seen in the glucodensities, their glucose is continuously fluctuating. The two-dimensional graphical representation of the density function of A1c and FPG helps to validate these findings.

\section{Discussion}

The primary contribution of this article is to propose a new representation of CGM data called glucodensity. We have validated this representation from a clinical point of view, proving that it is more accurate than time in range metrics.

\subsection{Diabetes etiology and biological components to capture in a mathematical represention}


Diabetes encompasses a heterogeneous group of impaired glucose metabolism,  such as the frequent presence of  hyperglycemias  or hypoglycemias  \cite{american20186}.  Anomalous glucose fluctuations are another essential trait of dysglycemic regulation \cite{Monnier1058,variabilidadmon}.  The use of glycemic control measures that go beyond the average glucose values such as A1c and also capture i) the impact of time spent at each glucose concentration on the glucose deregulation process, ii)  the oscillations of glucose associated with cellular damage \cite{Monnier1058}, is crucial in the management of  patients with diabetes as in the assessment of glucose metabolism with a high degree of precision.

\subsection{Clinical validation of glucodensity}

Our proposal accurately captures the components of diabetes mentioned above. Using clinical data, we evaluated the clinical sensitivity against established biomarkers in diabetes. We found  a high association between A1c, HOMA-IR, CONGA, MODD, MAGE, and glucodensity.  In the case of the HOMA-IR variable, the predictive ability does not seem excellent, although, to the best of our knowledge, no known marker shows a predictive ability against that variable. However, our model can provide consistent values in moderate and large HOMA-IR values.
While the fit for the variable A1c was not perfect, we must consider that the time scale for the A1c and the glucodensities were quite different.  A1c is a measure that reflects the average glucose over $2-4$ months while monitoring patients for less than $1$ week to compute the glucodensity. Our $R^2$ of $0.79$ is better than the average glucose recorded by the monitoring period ($R^2=0.61$), which indicates that an individual's glucose distributional values may give extra information to the long-term glucose averages.

In the prediction of glucodensity from A1c, HOMA-IR, and glycemic variability measures, the estimated $R^2$ shows a moderate relationship between those variables. However, we are introducing the essential variables of the glucose deregulation process. A possible explanation of this is that the use of the summary measures commonly used in diabetes can hardly capture an individual's glycemic profile. Glucose metabolism is very complex and highly dependent on the patient's conditions. For example, the cellular mechanisms are different in type I and type II diabetes. In the former, there is an inhibition of $\beta$-cell function and consequent non-insulin production, while insulin secretion is reduced in the latter \citep{Taylor1047}.  In this context, the introduction of the concept of glucodensity provides greater clinical accuracy to the possible decisions derived from such representation compared to traditional methods because we utilize the entire distribution of glucose concentrations of an individual over time.

\subsection{Time in range metrics vs. Glucodensity}

While time in range metrics may also achieve the previous aim, they do so to a clearly lesser extent than the glucodensity. Our proposal can capture the differences between individuals in each glucose concentration. Notwithstanding, time in range only measures glucose differences along intervals with the subsequent loss of information.   Also, time in range metrics are substantially limited since the target zones must be defined previously, and these may also depend on the study population or the aim of the analysis.

Empirical results demonstrate the advantages of our proposal out of the theoretical framework.   The ability of glucodensity to predict A1c, HOMA-IR, and the CONGA, MAGE, and MODD variability measures is surprisingly high, much higher than that achieved with the range metric despite using two different target zones: the deciles of normoglycemic patients glucose values and the target zones prescribed by the ADA. 

The estimated $R^2$ between glucodensities and A1c is similar than that reported by other authors between A1c and average glucose values \cite{media2}. However, in this study, patients are monitored only for $2-6$ days and not for weeks.  Two possible factors we must consider in the analysis of the results.  First, there are both diabetic and non-diabetic patients in our database, and, second, the glucodensity captures A1c better because it represents the entire distribution of glucose concentration values, while glycation rates are known to increase with glucose concentrations \cite{Singh2001}. In particular, the estimated $R^2$ between A1c and the mean glucose in our database is only $0.61$.

\subsection{Statistical considerations}

From a statistical standpoint, glucodensities are a special constrained type of functional data known as distributional data; therefore, it is not possible to directly use the usual statistical techniques. In this paper, we have proposed a framework for the analysis of these distributional data based on distances with existing techniques for hypothesis testing, cluster analysis, and regression models. However, further methodological development is necessary, as it can be the case of mixed models or causal inference methods where there is no available methodology.

\subsection{Limitations}

A potential limitation of our representation is that it ignores the order of events. Instead,  it analyzes only the distribution of glucose values.  However,  the event sequence may not be a  critical component in diabetes modeling.   The main factor of microvascular and macrovascular complications is chronic hyperglycemia \cite{Cryer2188,vskrha2016glucose}, and this is captured with high accuracy by our models. Moreover, an essential aspect of managing diabetes patients is hypoglycemia control, and our proposal also captures this. Finally, the third component of dysglycemia \cite{variabilidadmon}, glucose variability, can accurately predict by our representations, at least,  through metrics CONGA, MAGE, and MODD.

The sample size used may also be a limitation from a statistical point of view.  Nevertheless, in the field of diabetes, the AEGIS study is the world's largest databases and, unlike other studies, is composed of randomly selected individuals from a general population and non-participants \cite{zeevi2015personalized}. Finally, for study validation, perhaps the most reliable way of validating the new representation is in terms of the patients' long-term prognosis. However, to the best of our knowledge, no study with a reasonable sample size has this information from the intensive use of CGM technology.  Moreover, we have established the clinical validity from variables that do have a clear and established relationship with the prognosis and prevalence of diabetes as evidenced in the current literature in the field.

\subsection{Potential Applications}

Adopting the concept of glucodensity in clinical practice and biomedical research could be very promising in the following ways.

\begin{itemize}
    \item To have a simple and more accurate representation of the glycaemic profile of an individual. This representation is especially useful in the management of diabetic patients and to assess the effects of an intervention. 
    \item To establish if there are statistically significant differences between patients subjected to different interventions, for example, in a clinical trial.
    \item To identify different subtypes of patients based on their glycaemic condition and other variables. Cluster analysis of glucodensities can create new patient subtypes based on the risk of diabetes or other complications.  Furthermore, it allows us to describe the etiology of the disease better by creating groups of subjects whose glucose profiles and other clinical characteristics are similar.  
    \item To establish the prognosis or risk of a patient or to analyze the relationship of an individual's glycaemic profile with different clinical variables in epidemiological studies.
    \item To predict changes in the glycaemic profile based on the characteristics of the individuals and the intervention performed. For example: how does the glucodensity vary according to the diet?
    \item To recommend the most advantageous treatments for a patient. Following the previous idea, a causal inference model could be fitted where the response is glucodensity, for example, to establish which diet is the most beneficial for the individual to achieve a suitable glucose levels.
\end{itemize}

\subsection{Future work}

We introduce glucodensities methodology with CGM data. However,  our methodology is also valid for data from other biosensors such as accelerometers to measure physical activity levels.  In this domain, the time in range metric is one of the most used representations, and perhaps the adoption of our approach can lead to better results \cite{doi:10.1177/0962280217710835,Dumuid2020}.  The adoption of new methodology with other biosensors may be an essential research issue to be addressed in the future.

In the diabetes field, it will be necessary to evaluate the predictive capacity of the glucodensity in the long-term prognosis of patients. In addition, it would be interesting to assess, in more extended monitoring periods, the reproducibility between days and weeks with the representation constructed. One way to accomplish this is to compute the intraclass correlation coefficient (ICC) using, for example, the methodology proposed recently in \cite{doi:10.1111/biom.13287} and based on distances between functions.

\section*{Acknowledgements}

We thanks Russell Lyons for his discussions on the use of energy-distance based methods with glucodensities. 

 This work has received financial support from Carlos III Health Institute, Grant/Award Number: PI16/01395; Ministry of Economy and Competitiveness (SPAIN)  European Regional Development Fund (FEDER);  the Axencia Galega de Innovación, Consellería de Economía, Emprego e Industria, Xunta de Galicia, Spain, Grant/Award Number: GPC IN607B 2018/01; National Science Foundation, Grant/Award Number: DMS-1811888; the Spanish Ministry of Economy and Competitiveness Grant/Award Number: TIN2015-73566-JIN and TIN2017-84796-C21-R.

\section*{Competing Interests}
The authors declare no competing interests.

%
%

\bibliographystyle{ims}

\bibliography{referencias2.bib}

\end{document}